\documentclass[twocolumn,showpacs,preprintnumbers,amsmath,amssymb]{revtex4}
\usepackage{tabularx,graphicx}\begin{document}
%\documentstyle[aps]{revtex}
%\documentstyle[preprint,aps]{revtex}
%\begin{document}
\newcommand{\beq}{\begin{equation}}
\newcommand{\eeq}{\end{equation}}
\newcommand{\beqn}{\begin{eqnarray}}
\newcommand{\eeqn}{\end{eqnarray}}
\newcommand{\bmath}{\begin{subequations}}
\newcommand{\emath}{\end{subequations}}
%\draft
\title{Why holes are not like electrons. III. How  holes in the normal state turn into electrons in the superconducting state}
\author{J. E. Hirsch }
\address{Department of Physics, University of California, San Diego\\
La Jolla, CA 92093-0319}
 
\date{\today} 
\begin{abstract} 

In recent work, we discussed the difference between electrons and holes in energy band
in solids from a many-particle point of view, originating in  the electron-electron interaction\cite{hole1},
and from a single particle point of view, originating in  the electron-ion interaction\cite{hole2}. We proposed that
superconductivity in solids only occurs when the Fermi level is close to the top of a band (hole carriers), that it  originates in
`undressing' of carriers from $both$ the electron-electron and the electron-ion interaction,
and that as a consequence holes in the normal state behave like electrons in the superconducting state\cite{ijmp}.
However, the connection between both undressing effects was left unclear, as was left unclear how the
transformation from hole behavior to electron behavior occurs.
 Here we clarify these questions by showing that  the same electron-electron interaction physics that
promotes pairing of hole carriers and undressing of carriers from the electron-electron interaction 
leads to undressing of carriers from the electron-ion interaction and transforms the behavior of carriers from
hole-like to electron-like.   Furthermore this phenomenon is connected with the expulsion of negative charge
that we predict to occur in superconductors. These unexpected connections
support the validity of our theoretical framework, the theory of hole superconductivity, to explain superconductivity in solids. \end{abstract}
\maketitle 

\section{Introduction}

When Felix Bloch formulated his epoch-making theory of electrons in metals\cite{bloch1}, he had an ``uneasy feeling that the model of
independent electrons might represent a rather poor approximation and would turn out {\it in some respects} to be entirely inadequate''\cite{bloch2} (italic ours).
 Bloch realized full well the arbitrariness involved in his privileging the electron-ion interaction over the
direct electron-electron Coulomb interaction, given that $e^2=14.4 eV A$ is just as large an interaction between two electrons as between an electron and a monovalent ion at
the same distance\cite{bloch3}. In the statement quoted above, with ``in some respects'' Bloch had undoubtedly the phenomenon of superconductivity in mind\cite{bloch3}, which was widely believed at the time
to be caused by the electron-electron Coulomb interaction\cite{meissner}. However, Bloch's theory's myopic point of view became even more
myopic with  the conventional theory of superconductivity\cite{bcs}, that adscribes the phenomenon to the interaction of electrons with ions displaced from their equilibrium position
(electron-phonon interaction)\cite{frohlich}, bringing the disregard for the role of the electron-electron Coulomb interaction to a new high.

Instead, the theory of hole superconductivity\cite{holescth}
 proposes that while electron-electron Coulomb interactions can indeed be neglected when a band is almost empty, they become
increasingly dominant as the  filling  of a band increases. Unlike the prevalent point of view nowadays, that electron-electron interactions are most important near
half-filling of a band\cite{hubbard}, we propose that electron-electron interaction physics dominates and $qualitatively$ changes the normal metallic behavior
{\it when an electronic energy band is almost full},  in particular giving
rise to superconductivity. 

We propose that the normal metallic and the superconducting state are in a sense mirror images of each other, where the `mirror' switches the sign of the electric
charge, or   the bottom and top of the band, or  the electron-ion and the electron-electron interactions. The `mirror' is by no means perfect because 
of the vastly different masses associated with 
the negative electron and the positive ion, hence there are significant differences between the normal metallic and the superconducting state, and the `mirror switching' does not occur
right at the half filled band but instead only when a band is almost full.
Nevertheless, an essential commonality between both states emerges from considering the electron-ion and the electron-electron interaction on an equal footing, leading to the
conclusion that normal metallic behavior ensues when the electron-ion interaction
dominates, and superconducting behavior when the electron-electron interaction dominates.

 In the normal metal, 
there are few electrons  in the band with a lot of room to move around, the positive ions are rigid and the electrons will act as nearly independent of
each other and adjust their wavefunctions $individually$ to optimize the electron-ion interaction, with the
electron-electron interaction being non-optimized. In the superconductor instead, the almost-full band is crowded with a lot of
electrons that become 'rigid' because of their mutual strong interaction, with  their wavefunction adjusted to optimize the electron-electron
interaction and with the electron-ion interaction being non-optimized. This means in particular that states that are
electron-ion-interaction-energy costly will be occupied in the superconducting state.   
A consequence of this point of view is also that the theory  predicts that superconductivity is particularly favored when the electron-ion interaction strength $Ze^2$ is weak, i.e. when 
 the ionic charge $Z$ is small\cite{diat1}.

In the following, we analyze that aspect  of the Coulomb interaction that we believe to be essential to understand superconductivity, at the level of a single atom,
a diatomic molecule and a solid. We show that the essential physics manifests itself in both real and momentum space in a remarkably parallel fashion
in going from the atom to the solid. It involves $expansion$ of the electronic wavefunction to achieve lowering of electron-electron interaction and of
quantum kinetic energy, at the expense of electron-ion interaction energy. 

 \section{the atom}
 
 Consider the wavefunction of an electron in the lowest energy state (1s) of a hydrogen-like atom of nuclear charge $Z$
 \beq
 \varphi_Z(r)=(\frac{Z^3}{\pi})^{1/2}e^{-Zr}
 \eeq
 with $r$ measured in units of the Bohr radius $a_0$. For two electrons in the atom in a singlet state, the spatial wave function is $not$
 \beq
 \psi(r_1,r_2)=\varphi_Z(r_1)\varphi_Z(r_2)
 \eeq
 because the strong Coulomb interaction between two electrons in this state
 \beq
 U=\int d^3r d^3r' |\varphi_Z(r_1)|^2 \frac{e^2}{|\vec{r}_1-\vec{r}_2|}|\varphi_Z(r_2)|^2
 \eeq
 makes this state energetically too costly. The true wave function in the two-electron atom contains both radial and angular correlations between the
 electrons, and is well described by the Hylleraas wave function\cite{hylleraas}. However, the main effect is captured by the Hartree approximation that allows for
 expansion of the single-particle orbital:
 \bmath
 \beq
 \psi_H(r_1,r_2)=\varphi_{\bar{Z}}(r_1)\varphi_{\bar{Z}}(r_2)
 \eeq
 with 
 \beq
 \bar{Z}=Z-y
 \eeq
 \emath
 The energy of the electrons in that state is (expressed in Rydbergs=13.6eV)
 \beq
 E(\bar{Z})=E_{kin}(\bar{Z})+E_{e-i}(\bar{Z})+E_{e-e}(\bar{Z})
 \eeq
 with
 \bmath
 \beq
 E_{kin}(\bar{Z})=2\bar{Z}^2=E_{kin}(Z)-4Zy+2y^2
 \eeq
 \beq
 E_{e-i}(\bar{Z})=-4Z \bar{Z}=E_{e-i}(Z)+4Zy 
  \eeq
 \beq
 E_{e-e}(\bar{Z}) =\frac{5}{4}\bar{Z} =E_{e-e}(Z)-\frac{5}{4}y
 \eeq
 \emath
 with $E_{e-e}(Z)=U$. Therefore, expansion of the orbital $costs$ electron-ion energy
 \bmath
 \beq
 \Delta E_{e-i}=4Zy
 \eeq
but gives a lowering of both kinetic energy
  \beq
 \Delta E_{kin}=-4Zy+2y^2
  \eeq
 and of Coulomb energy
  \beq
 \Delta E_{e-e}=-\frac{5}{4} y . \eeq 
 \emath
 The single-particle energy of each electron in the expanded orbital is the sum of its kinetic and electron-ion energy, and is $larger$ than the
 single-particle energy in the unexpanded orbital by
 \beq
 \epsilon=\frac{1}{2}(\Delta E_{kin}+ \Delta E_{e-i})=y^2
 \eeq
 because the cost in electron-ion energy is larger than the kinetic-energy lowering. However, this is more than offset by the reduction in Coulomb  repulsion in the expanded orbital:
\beq
U'\equiv E_{e-e}(\bar{Z})=U-\frac{5}{4} y
\eeq
so that the total energy 
\beq
U'+2\epsilon=U+2y^2-\frac{5}{4} y
\eeq
is lower than the energy in the non-expanded doubly occupied orbital provided   $y<5/8$.  
  \begin{figure}
\resizebox{8.5cm}{!}{\includegraphics[width=7cm]{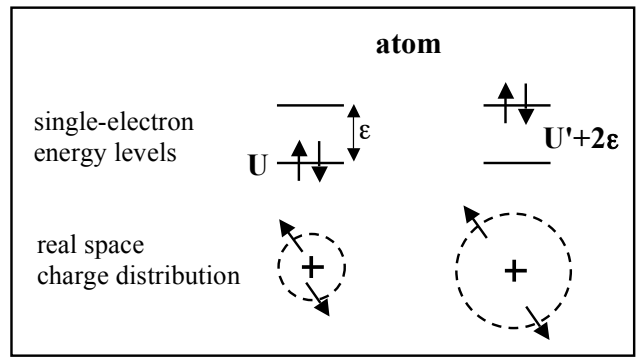}}
  \caption{The two-electron atom. When the two electrons occupy the unexpanded orbital, they pay a high price in Coulomb energy, $U$. Instead, they prefer to reside in an 
  expanded orbital, paying each the price $\epsilon$ in single-particle energy, yet achieving a lower total energy $U'+2\epsilon$.  }
\end{figure}

 We represent the situation schematically in Fig. 1, with two ``orbitals'' per atom. The two electrons ``choose'' to occupy the orbital with
higher single-particle energy because the lower Coulomb repulsion more than offsets the cost in single-particle energy. The space-charge
distribution becomes more extended in real space: negative charge is ``expelled'' outward when the second electron is added to the
orbital, leading to the lowering of both electron-electron Coulomb energy (Eq. 7c)) and kinetic energy (Eq. (7b)).
 
 The optimal value of $y$ that minimizes the energy Eq. (10) is $y=5/16$, so the orbital Eq. (1) expands to $\varphi_{\bar{Z}}(r)$ with
$\bar{Z}=Z-\frac{5}{16}$,
which costs electron-ion Coulomb energy
$ \Delta E_{e-i}=\frac{5}{4}Z$
 but gives a lowering of kinetic energy
 $\Delta E_{kin}=-\frac{5}{4}Z+\frac{25}{128}$
 and of electron-electron Coulomb energy
 $\Delta E_{e-e}=-\frac{25}{64}$. 
 The single-particle energy of each electron in the expanded orbital is larger than the
 single-particle energy in the unexpanded orbital by
 $\epsilon= \frac{25}{256}$
which is more than offset by the reduction in Coulomb  repulsion in the expanded orbital
$U'=U(Z-5/16)=U(Z)-\frac{25}{64}$,
so that the total energy is
$U'+2\epsilon=U-\frac{25}{128}$
and the energy lowering achieved by expanding the orbital is  $25/128Ry=2.7 eV$.

The physics of the two-electron atom just described is a remarkable microcosm of the physics of electrons in a nearly filled electronic energy band: orbital expansion and
promotion to higher single-particle  energy levels driven by electron-electron repulsion and kinetic energy lowering will play a key role, as we discuss in the
following sections.

\section{the diatomic molecule}

Within a linear combination of atomic orbitals approach, the bonding and antibonding orbitals of a diatomic molecule are given by
\beq
\psi_{b,a}(r)=\frac{\varphi_1(r) \pm  \varphi_2(r)}{(2(1\pm  S_{12}))^{1/2}}
\eeq
with the $+$ ($-$) sign corresponding to bonding (b)  (antibonding (a)), assuming s-orbitals for definiteness. $S_{12}=(\varphi_1,\varphi_2)$ is the overlap
matrix element. The bonding orbital has larger amplitude and correspondingly larger charge density in the region between 
the atoms, while the antibonding orbital changes sign and has vanishing charge density at a point between the atoms, as depicted in Fig. 2(a). 
As emphasized in II, this is a real
physical difference between bonding and antibonding orbitals that cannot be eliminated by a canonical transformation.

  \begin{figure}
\resizebox{8.0cm}{!}{\includegraphics[width=7cm]{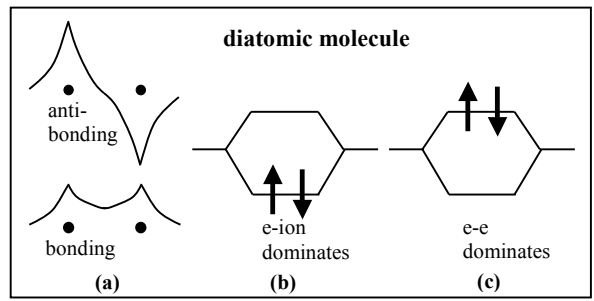}}
  \caption{(a) Schematic depiction of the wavefunction for bonding and antibonding orbitals in a diatomic molecule. The bonding orbital has higher charge density between the ions to
  get maximum benefit from the electron-ion interaction and a smooth wavefunction to give lower kinetic energy. (b) and (c) shows the bonding (lower) and 
  antibonding (upper) single-particle energy levels, with different occupations of two electrons of opposite spin.  }
\end{figure}

Putting two electrons in the bonding orbital results in an electron-electron repulsion energy
\bmath
\beqn
U_{bb}&=&\int d^3r d^3r' |\psi_b(r)|^2\frac{e^2}{|\vec{r}-\vec{r'}|} |\psi_b(r')|^2 \nonumber \\
&=& \frac{U+V+2J+4\Delta t}{2(1+S_{12}))^2}
\eeqn
 and for two electrons in the antibonding orbital 
\beqn
U_{aa}=\frac{U+V+2J-4\Delta t}{2(1-S_{12}))^2}
\eeqn
\emath
where, in terms of the Coulomb integrals
\beq
(ij|kl)\equiv \int d^3r d^3r' \varphi_i^*(r)\varphi_j^*(r')\frac{e^2}{|\vec{r}-\vec{r}'|} \varphi_l(r')\varphi_k(r)
\eeq
$U=(ii|ii)$, $V=(ij|ij)$, $J=(ij|ji)=(ii|jj)$ and $\Delta t=(ii|ij)$.
All these matrix elements are positive. In particular,  the hybrid Coulomb matrix element
\beq
\Delta t=(ii|ij)
\eeq
lowers the Coulomb repulsion for two electrons in antibonding states\cite{bondcharge}, as seen from Eq. (12b). For sufficiently large $\Delta t$, the  `inverted occupation'
shown in Fig. 2(c) with the two electrons in the antibonding orbital would have lower total energy 
(kinetic + electron-ion + electron-electron) than the usual one where both electrons are in the bonding orbital, Fig. 2(b).

Note also that for one electron in the bonding state and one in the antibonding state, the direct Coulomb repulsion is
\beq
U_{ba}=\frac{U+V-2J}{2(1-S_{12}^2)}
\eeq
which indicates that the Coulomb matrix element $J$ favors ferromagnetism, particularly near the half-filled band\cite{fm}.

Let us now consider the diatomic molecule as a microcosm for an electronic energy band, as depicted in Fig. 3. If the electron-ion
energy dominates, the occupation will be as shown on the left side. In particular, for three electrons in the molecule
(analogous to a `nearly filled band') two electrons will go into the bonding state and one electron into the antibonding state.
Instead, if the electron-electron interaction dominates, the occupation will be as depicted on the right side of Fig. 3, where two
electrons occupy the antibonding state and one electron the bonding state. Equivalently,
{\it the single hole in the filled band resides in the bonding rather than in the antibonding state}.

  \begin{figure}
\resizebox{8.0cm}{!}{\includegraphics[width=7cm]{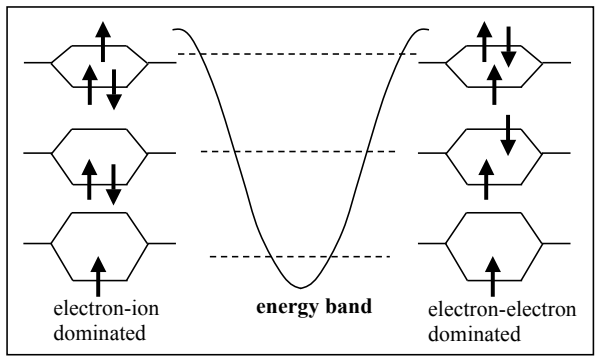}}
  \caption{The diatomic molecule as a microcosm for an energy band. When the electron-electron interaction dominates,
  the occupation of single-particle energy levels changes, as shown on the right side of the figure:
the middle diagram represents a ferromagnet and the upper diagram a superconductor.}
\end{figure}

In a tight binding description, the Hamiltonian for electrons in a diatomic molecule with one orbital per atom is given by\cite{chemphys}
\beqn
H&=&-\sum_\sigma  [t_0 - \Delta t (n_{1,- \sigma} +n_{2,- \sigma})]+U\sum_in_{i\uparrow}n_{i\downarrow} \nonumber \\
& &+V\sum_\sigma n_{1\sigma} n_{2\sigma} + J(\sum_\sigma(c_{1\sigma}^\dagger c_{2\sigma}+h.c.))^2
\eeqn
assuming higher order interactions (involving $6$ or more fermion operators) can be neglected. 
$t_0$ is the hopping amplitude for a single  electron in the molecule. The creation and annihilation operators
in 
Eq. (16) refer to {\it orthogonal orbitals} $\phi_1(r), \phi_2(r)$ obtained from linear combinations of the atomic orbitals
$\varphi_1(r), \varphi_2(r)$\cite{chemphys}. The Coulomb matrix elements are given by the expression Eq. (13) with the orthogonal orbitals.
In particular, the hybrid matrix element $\Delta t$ is given in terms of  Eq. (13) with atomic orbitals, $\Delta t_{a.o.}$, by
\beq
\Delta t=\Delta t_{a.o.}-\frac{S_{12}}{2}(U+V)
\eeq
to lowest order in the nearest neighbor overlap $S_{12}$. To the extent that the Mulliken approximation for the overlap charge distribution holds\cite{mulliken}, the expression
Eq. (17) exactly vanishes, and in practice it is found that deviations from the Mulliken approximation are very small\cite{chemphys}. For that reason,
the "correlated hopping" interaction term $\Delta t$ is usually ignored in formulating tight binding Hamiltonians to describe interacting
electrons in solids.

However, in considering the interactions between electrons as the number of electrons in the diatomic molecule (or the band) increases it is
essential to take into account the expansion of the atomic orbital that occurs for the doubly-occupied atom, discussed in Sect. II. 
When doing so, we showed in Refs. \cite{diat1,diat2,diat3} that the correlated hopping term $\Delta t$ has a value very different from that
given by Eq. (17): it is increasingly positive as the interatomic distance $R$ decreases and   the ionic charge $Z$ decreases.
The `correlated hopping' parameter $\Delta t$ changes the hopping amplitude depending on the electronic occupation of the sites involved in
the hopping process: the hopping amplitude for an electron is $t_0$, $t_0-\Delta t$ and $t_0-2\Delta t$ depending on whether there are $0$, $1$ or $2$ other
electrons at the sites involved in the hopping process.

For one electron in the diatomic molecule, the eigenstates of the Hamiltonian Eq. (16) are the bonding and antibonding states $c_{b\sigma}^\dagger|0>, c_{a\sigma}^\dagger|0>$, with
$|0>$ the empty molecule and 
\bmath
\beq
c_{b\sigma}^\dagger=\frac{c_{1\sigma}^\dagger+c_{2\sigma}^\dagger}{\sqrt{2}}
\eeq
\beq
c_{a\sigma}^\dagger=\frac{c_{1\sigma}^\dagger-c_{2\sigma}^\dagger}{\sqrt{2}}
\eeq
\emath
and energies $\epsilon_b=-t_0$, $\epsilon_a=+t_0$. For three electrons in the molecule, the eigenstates are
\bmath
\beq
|\tilde{b}>_\sigma=c_{a\sigma}^\dagger c_{b\uparrow}^\dagger  c_{b\downarrow}^\dagger |0>
\eeq
\beq
|\tilde{a}>_\sigma=c_{a\uparrow}^\dagger  c_{a\downarrow}^\dagger  c_{b\sigma}^\dagger |0>
\eeq
\emath
with energies
\bmath
\beq
\tilde{\epsilon}_b=-(t_0-2\Delta t)+U+2V
\eeq
\beq
\tilde{\epsilon}_a=+(t_0-2\Delta t)+U+2V
\eeq
\emath
respectively. The ordering of these states depends on the magnitude of the single hole hopping amplitude
\beq
t_h\equiv t_0-2\Delta t
\eeq
For $t_h>0$, $\tilde{\epsilon}_b < \tilde{\epsilon}_a$ and the state $|\tilde{b}>_\sigma$, corresponding to two electrons in the bonding orbital and one electron in the antibonding orbital has lower energy than the
state $|\tilde{a}>_\sigma$, so the lowest energy state corresponds to the three-electron state on the left side of Fig. 3. Instead, for $t_h<0$ the ordering is reversed and two electrons will occupy the
antibonding orbital, as depicted on the right side of Fig. 3. The latter situation will occur if $\Delta t>t_0/2$.

The parameter $\Delta t$ represents a `bond charge repulsion'\cite{bcr} between electrons: it pushes electrons away from the bond (region between the  ions)  towards the sites (ions) by suppressing
the occupation of the bonding state and increasing the occupation of the antibonding state.   The magnitude of the hopping parameter $t_0$, which is determined   by the strength of the
electron-ion attraction, decreases as the ionic charge $Z$ decreases. Instead, as already mentioned, the magnitude of $\Delta t$  $increases$ as the ionic charge decreases
because of the increased orbital expansion under double occupancy. So when the electron-electron interaction dominates over the electron-ion attraction in the diatomic molecule, 
$t_h$ changes sign and this leads to the inverted occupation  shown on the
right side of Fig. 3.

Furthermore, as the electrons doubly-occupy the antibonding orbital, they have larger amplitude for their wavefunction at the atomic site,
rather than in the interatomic bond, as depicted by the upper picture in Fig. 2(a). Larger on-site occupation implies a larger expansion of the
atomic wavefunction, as depicted in the lower right panel of Fig. 1, and hence negative charge is expelled $outward$, away from the
interatomic region. This in turn also causes a relative lowering of kinetic energy, as discussed in Sect. II for the single atom.

In summary, our analysis of the diatomic molecule shows strong parallels with the discussion in Sect. II for the single atom: when the electron-electron
interaction dominates over the electron-ion interaction, the single-electron higher energy level (antibonding state)  becomes more occupied and the single-electron lower energy level
(bonding state) becomes less occupied.
In other words, holes tend to increasingly occupy the lower-energy single-electron energy levels. At the same time, the  negative
charge spatial distribution increases $outward$, which implies a higher positive charge distribution in the region inside the structure.
That is, negative charge is ``expelled'' from the interior region of the molecule towards the exterior.
The interaction matrix element responsible for this effect in the diatomic molecule (and in the solid as we will see))  is $\Delta t$, and the essential underlying physics is
orbital expansion driven by electron-electron interaction
(since, as mentioned, in the absence of orbital expansion the parameter $\Delta t$ is found to be zero) and 
kinetic energy lowering (since the atomic orbital expansion gives kinetic energy lowering as seen in Sect. I).

\section{Further analysis  of $\Delta t$}
In this section we discuss in more detail the physical origin of the occupation inversion that we argue can occur in the diatomic molecule, because the same principles
will apply to the solid state. 
In Refs. \cite{diat1,diat2,diat3} we calculated from first principles the hopping amplitude for a single electron, $t_0$, and for a single hole, $t_h$, in
a diatomic molecule, taking into account the orbital relaxation effect. These hopping amplitudes were obtained from the difference in 
energy of an electron or a hole in the bonding and antibonding states. As can be seen for example in Fig. 7(c) of Ref. \cite{diat3}, the hole hopping amplitude $t_h$ (denoted by $t_2$ in ref.\cite{diat3}) goes to zero
{\it and in fact changes sign} for sufficiently small interatomic distance $R$ and small ionic charge $Z$. In this section we discuss in more detail how
the change in sign of $t_h$ comes about.

In paper I of this series\cite{hole1} we have emphasized the change in hopping amplitude with occupation that arises from the
overlap matrix element of the expanded and unexpanded atomic orbital. For the $1s$ orbital discussed in sect. II, the atomic overlap matrix element is given by
\beq
S=(\varphi_Z,\varphi_{\bar{Z}})=\frac{(Z\bar{Z})^{3/2}}{(\frac{Z+\bar{Z}}{2})^3}
\eeq
with $\bar{Z}=Z-5/16$. $S$ is a decreasing function of the ionic charge $Z$, and as $Z\rightarrow 5/16$ it becomes arbitrarily small. The hopping amplitude
for an electron when there are $m$ other electrons in the two sites involved in the hopping process is $t_m=S^m t_0$, and has the same sign as $t_0$. If we define
$\Delta t$ as the difference in hopping amplitude for a single hole and when there is a second hole in one of the sites, it is given by
\beq
\Delta t=t_0S(1-S)
\eeq
This physics will promote pairing of hole carriers\cite{holesc}, but will $not$ change the sign of the hopping amplitude for holes, since for a single hole we have
\beq
t_h=t_0S^2
\eeq
of the same sign as $t_0$.

However, there are additional contributions to the variation of hopping amplitude with occupation, that arise from matrix elements of the Coulomb
interaction involving electrons on neighboring sites. In the absence of orbital expansion these contributions essentially cancel, as discussed in Sect. III, however
in combination with orbital expansion they have a fundamental effect, and Eq. (24) ceases to be correct. We discussed in Ref.\cite{diat1}
a calculation of the hopping amplitude for a single hole in the diatomic molecule obtained by approximating the bonding and antibonding states of the
three-electron diatomic molecule by the wavefunction
\beqn
& &\tilde{\psi}_{b,a}(r_1,r_2,r_3)= \nonumber \\
& &\frac{\bar{\varphi}_1(r_1)      \bar{\varphi}_1(r_2)     \varphi_2(r_3)    
\pm  \varphi_1(r_1)    \bar{\varphi}_2(r_2)      \bar{\varphi}_2(r_3) 
 }{(2(1\pm \tilde{S}_{12}))^{1/2}}
\eeqn
where $\bar{\varphi}_i(r_j)$ and $ \varphi_i(r_j)$ are the expanded and unexpanded orbitals for the $j-th$ electron in atom $i=1,2$ and the overlap is
\beq
\tilde{S}_{12}=(\bar{\varphi}_1,\varphi_1)^2(\bar{\varphi}_1,\varphi_2).
\eeq
where $(\bar{\varphi}_1,\varphi_1)=S$ ((Eq. (22)). The energies of these states are
\beq
\tilde{\epsilon}_{b,a}=\frac{(\bar{1}\bar{1} 2|H_3|\bar{1}\bar{1} 2) \pm (\bar{1}\bar{1} 2|H_3|1\bar{2}\bar{2})}{1\pm   \tilde{S}_{12}}
\eeq
where
 the labels $1, \bar{1}$ are shorthand
for the orbitals $\varphi_1,\bar{\varphi}_1$, etc.
The three-electron Hamiltonian is given by
\bmath
\beq
H_3=H_{3sp}+H_{3ee}
\eeq
with the single-particle and electron-electron parts given by
\beq
H_{3sp}=h_1+h_2+h_3
\eeq
\beq
H_{3ee}=h_{12}+h_{23}+h_{13}
\eeq
\emath
where $h_i$ is the single-particle Hamiltonian for the $i$-th electron and $h_{ij}$ the interaction potential between the $i$-th and $j$-th electron, given by (in atomic units)
\bmath
\beq
h=-\nabla^2-\frac{2Z}{r_1}-\frac{2Z}{r_2}
\eeq
\beq
h_{12}=\frac{2}{r_{12}}
\eeq
\emath
with $r_1, r_2$ the electronic coordinate relative to atom $1$, $2$, and $r_{12}$ the distance between electrons. The difference between the energies Eq. (27) give the hole hopping 
amplitude $t_h$
\beq
t_h=\frac{\tilde{\epsilon}_a-\tilde{\epsilon}_b}{2}.
\eeq
In the regime where the overlap $S$ is small (small ionic charge $Z$) we can neglect the overlap Eq. (26), being proportional to $S^2$. The difference in energies between the 
bonding and antibonding states is then determined by the second matrix element in Eq. (27), given by
\beq
(\bar{1}\bar{1} 2|H_3|1\bar{2}\bar{2})=(\bar{1}\bar{1} 2|H_{3sp}|1\bar{2}\bar{2})+(\bar{1}\bar{1} 2|H_{3ee}|1\bar{2}\bar{2})
\eeq
The first term in Eq. (31) (single-particle term) is
\beq
(\bar{1}\bar{1} 2|H_{3sp}|1\bar{2}\bar{2})=(\bar{1},h,\bar{2}) S^2+2(\bar{1},h,\bar{1})SS_{\bar{1}\bar{2}} 
\eeq
and is negative, so it lowers the energy of the bonding state and raises the energy of the antibonding state.
The first term in Eq. (32) corresponds essentially to Eq. (24), the second term would be absent if the orbitals at neighboring sites were constructed to be
orthogonal to each other.

The second term in Eq. (31) (electron-electron interaction term) is
\beq
(\bar{1}\bar{1} 2|H_{3ee}|1\bar{2}\bar{2})=2(\bar{1},\bar{1},h_{12},1,\bar{2})S 
+(\bar{1},2,h_{12},1,\bar{2})  S_{\bar{1}\bar{2}}
\eeq
and is positive, so it raises the energy of the bonding state and lowers the energy of  the antibonding state.
It is clear then that the state of inverted occupation (antibonding state) has lower electron-electron interaction energy than the ordinary state (bonding state).
The difference in energy between the states is the hole hopping amplitude
\beqn
t_h&=&-(\bar{1},h,\bar{2}) S^2-2(\bar{1},h,\bar{1})S S_{\bar{1}\bar{2}}  \nonumber \\
&-& 2(\bar{1},\bar{1},h_{12},1,\bar{2})S  
-(\bar{1},2,h_{12},1,\bar{2})  S_{\bar{1}\bar{2}}
\eeqn
where the first two terms, involving the electron-ion interaction, are positive (including their sign), and the last two terms involving the electron-electron interaction are  negative (including their sign) .
When the electron-electron interaction dominates over the electron-ion interaction (for small ionic charge $Z$) the sign of $t_h$ switches from positive to negative,
and the state of inverted occupation becomes the low-energy state.

In Ref.\cite{diat1} we only explored the parameter regime where $t_h$ has the same sign as $t_0$, i.e. where there is no occupation inversion. For
sufficiently small $Z$ and small interatomic distance $R$, $t_h$ is found to change sign, as shown in Fig. 4. (For the calculation in Fig. 4 the overlap Eq. (26) was included 
hence there are other terms in $t_h$ beyond those given in Eq. (34), see Ref.\cite{diat1}). The region $t_h<0$ corresponds to the electron-electron
dominated regime shown on the right side of Fig. 3, where occupation inversion occurs.
For further details on the form of the matrix elements entering in the calculation leading to Fig. 4 
the reader is referred to the Appendix of Ref.\cite{diat1}.

  \begin{figure}
\resizebox{8.5cm}{!}{\includegraphics[width=7cm]{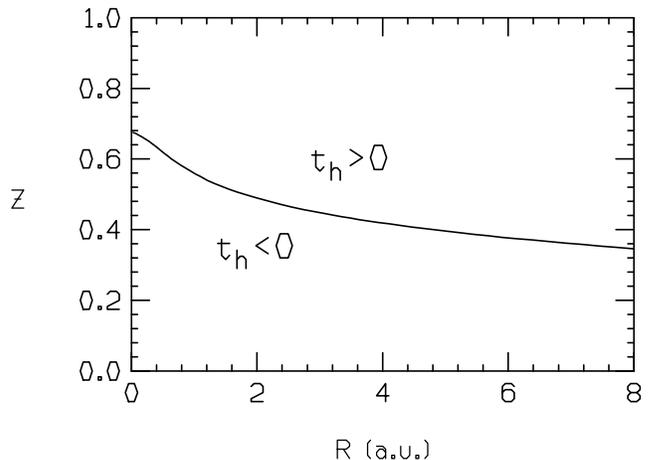}}
  \caption{ Dependence of the   sign of the single hole hopping amplitude in the diatomic molecule $t_h$ on the ionic charge $Z$ and the interatomic distance $R$ for 
  the $1s$ orbital.
 In the region above the curve the single hole hopping amplitude has the same sign as the single electron hopping amplitude $t_0$ ($t_0>0$), corresponding to
 the ordering of energy levels shown on the left side of Fig. 3. In the region below the curve the sign of the single hole hopping amplitude is opposite to that
 of the single electron hopping amplitude and the occupation of the levels for three electrons in the diatomic molecule is inverted as shown on the
 right side of Fig. 3. }
\end{figure}

\section{the solid}
We consider the Hamiltonian for electrons in a band in a tight binding representation given by\cite{bondcharge}
\beqn
H&=&-\sum_{<ij>,\sigma} (t_0-\Delta t (n_{i,-\sigma}+n_{j,-\sigma}))(c_{i\sigma}^\dagger c_{j\sigma}+h.c.) \nonumber \\
& & +U\sum_i n_{i\uparrow} n_{i\downarrow}+V\sum_{<ij>} n_i n_j
\eeqn
Performing  a particle hole transformation
\beq
h_{i\sigma}^\dagger=c_{i\sigma}
\eeq
the Hamiltonian becomes
\beqn
H&=&+\sum_{<ij>,\sigma} (t_h+\Delta t (n_{i,-\sigma}+n_{j,-\sigma}))(h_{i\sigma}^\dagger h_{j\sigma}+h.c.) \nonumber \\
& & +U\sum_i n_{i\uparrow} n_{i\downarrow}+V\sum_{<ij>} n_i n_j
\eeqn
where the number operators are now $n_{i\sigma}=h_{i\sigma}^\dagger h_{i\sigma}$ and 
\beq
t_h=t_0-2\Delta t   .
\eeq
We have shown that this Hamiltonian leads to pairing of holes and to superconductivity in a BCS formalism in the regime of low
hole concentration\cite{holesc}. Furthermore we have discussed that the transition to superconductivity driven by $\Delta t$ is associated with
``undressing''\cite{undr}: the effective mass of the carriers decreases and the quasiparticle weight increases, and a transfer of spectral weight
from high to low frequencies takes place both in the single particle spectral function (detectable in photoemission experiments) and in the
optical absorption spectrum\cite{apparent}.

In addition, in paper II of this series we have argued that superconductivity is also associated with `undressing' of carriers from the
electron-ion interaction, and that it involves a {\it wavelength expansion}: the wavelength of carriers at the Fermi energy
grows from a microscopic length ($k_F^{-1}\sim \AA^{-1}\sim$ interatomic distance) to a much larger wavelength. Here we explain how this occurs.

  \begin{figure}
\resizebox{8.5cm}{!}{\includegraphics[width=7cm]{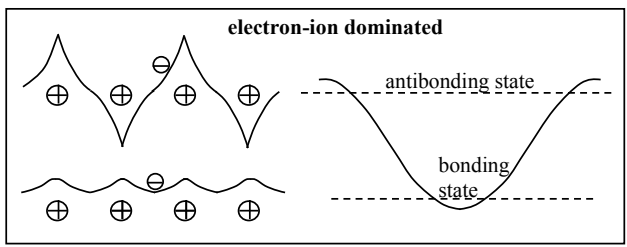}}
  \caption{The electronic states at the Fermi energy look very different when the Fermi level is near the bottom and near the top of a band in a normal metal.
  Near the bottom of the band the wavelength is large and the state is smooth, resembling the free-electron plane wave function, nearly unaffected by the
  electron-ion potential. Instead, near the top of the band the wavelength approaches a single lattice spacing and the state is strongly modified
  by the electron-ion interaction.}
\end{figure}

Figure 5 shows schematically the form of the wavefunction of the carriers at the Fermi energy when the band is almost empty and when it is almost full.
We are assuming that the hopping amplitude $t_0$ is positive, as occurs for $s$ orbitals, which implies that the minimum in the band
occurs for $k=0$. 
When the Fermi level is close to the top of the band (hole carriers) the state is ``bumpy'' rather than smooth. As discussed in II, it is not a state well-suited
to conduct electricity: it is strongly coupled to the lattice and is (nearly) Bragg-scattered by it, transferring the momentum it acquires from an external field
to the ionic lattice in the process, and acquiring momentum in opposite direction. It is highly ``dressed'' by the electron-ion interaction. In addition, 
as discussed in I, as the hole propagates it causes a large deformation in the atomic cloud where it lands (Fig. 1), and as a consequence it is also highly dressed by the
electron-electron interaction.

As the system goes superconducting the carriers at the Fermi energy undress from both the electron-electron\cite{hole1}  and the electron-ion interaction\cite{hole2}
 in order to conduct
better. Their wavelength expands, and they no longer `see' the discrete electron-ion potential\cite{hole2}.     {\it Figure 6 illustrates how this occurs}.

  \begin{figure}
\resizebox{8.5cm}{!}{\includegraphics[width=7cm]{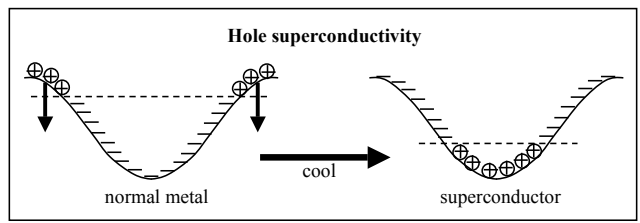}}
  \caption{When the system goes superconducting,  the antibonding states near the top of the band become fully occupied 
  with electrons and the 
  bonding states at the bottom of the band become empty. In other words, the holes  ``Bose-condense'' to the bottom of the band. The promotion
  of electrons to the higher electron-ion energy states is driven principally by the electron-electron interaction. }
\end{figure}

Indeed, we propose that in the transition to superconductivity, the electronic occupation of the band is shifted upward, to occupy all the single-particle energy
levels up to the top of the band, leaving the lowest energy levels in the band empty. In other words, the holes that were occupying the uppermost 
electronic levels ``Bose-condense'' and now occupy the lowest electron-ion energy levels. By so doing, the electronic states that are now at the boundary between 
occupied and empty states (dashed line on the right side of Fig. 6) become ``smooth'' states, of the 'bonding type', with large amplitude in the region between the ions\cite{mrs}.

How can such  complete reorganization take place? Just like in the diatomic molecule, it will occur for low hole concentration if the hopping amplitude $t(n)$ 
($n$=number of electrons per site) changes sign:
\beq
t(n)=t_0-n\Delta t <0
\eeq
so that in particular the single hole hopping amplitude $t_h=t_0-2\Delta t <0$. Then, in the Hamiltonian Eq. (37) the lowest energy state for a hole
occurs for $k=0$ rather than for $k=\pi$. This, however, is not just a `canonical transformation':  the states near the bottom and the top of
the band are $qualitatively$ different.

  \begin{figure}
\resizebox{8.5cm}{!}{\includegraphics[width=7cm]{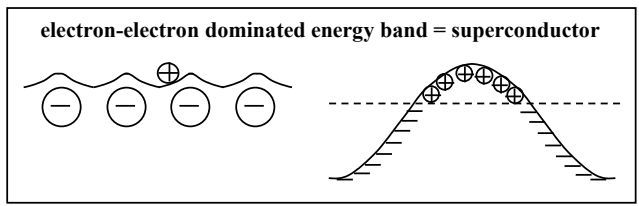}}
  \caption{ In the superconducting state, the holes occupy smooth bonding states, that were previously occupied by electrons in the normal state.
  The holes `ride' on top of the rigid negatively charged ions without disturbing their charge distribution as they do in the normal state\cite{mrs}.
  The $electron$ energy band depicted on the right side includes the   electron-electron interaction energy $n\Delta t$ and therefore it is inverted.}
\end{figure}

As function of temperature, the hopping amplitude $t(n)$ can change sign if the number of electrons $n$ increases as the temperature is lowered.
Indeed we showed in Ref. \cite{normal} that several anomalous properties of high $T_c$ cuprates are explained by the assumption that the number of holes in the $Cu-O$ planes 
decreases as the temperature is lowered, and that such a charge transfer process between the planes and off-plane charge reservoir atoms is aided by the Coulomb matrix
element $\Delta t$. In other systems the change in $n$ with temperature may be associated with redistribution of occupation between different bands without charge transfer between different 
regions in real space.

In Figure 7 we represent the situation in an alternative way, where the 'energy band' now includes the electron-electron interaction energy
$n\Delta t$. Because the sign of the hopping amplitude has changed for the almost-full band, the lowest energy electronic states occur at $k=\pi$ and 
are occupied by electrons, and the highest occur 
at $k=0$ and are unoccupied. Because the band is almost full, 
the carriers at the Fermi energy now have a smooth wavefunction, as depicted on the left side of Fig. 7, just like the smooth
wavefunction of carriers at the Fermi energy in the normal state of a metal when the band is almost empty.
The holes ride `on top' of the negative ions, without disturbing them and without noticing the discreteness of the potential. This corresponds to the
`mirror image' of the lower left panel of Fig. 5.

This reorganization in energy level occupation has a counterpart in the real space charge distribution. Figure 8 shows the situation schematically for the
atom, the diatomic molecule and the solid. As higher single-electron energy levels become occupied, electronic charge moves $outward$
in the three cases.
The theory of hole superconductivity predicts that negative charge is expelled from the interior of the metal towards the surface
as it undergoes the transition to the superconducting state\cite{chargeexp}, and that as a consequence an excess negative charge density $\rho_-$ exists within a 
London penetration depth of the surface of superconductors, with $\rho_-=-H_c/4\pi$ or $\rho_-=-H_{c1}/4\pi$ for type I and type II
superconductors respectively\cite{electrospin}.

The superconductor looks like a `giant atom'\cite{giantatom} with expanded orbitals.

  \begin{figure}
\resizebox{8.5cm}{!}{\includegraphics[width=7cm]{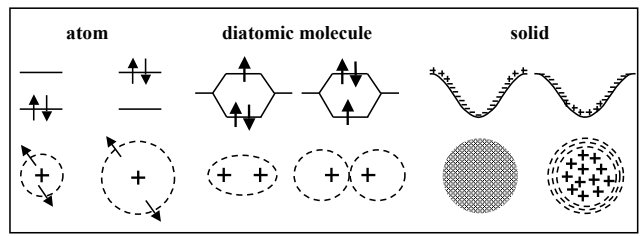}}
  \caption{Single-electron energy levels (upper part of the figure) and real space charge distribution (lower part of the figure).
  As electrons   move up the energy levels, negative charge moves $outward$ in the atom, the diatomic molecule and
  the solid in the superconducting state. In the solid in the normal state, the charge distribution is spatially homogeneous.
  The charge redistribution is driven by electron-electron repulsion and kinetic energy lowering, and is larger for smaller    ionic charge $Z$
  in all three cases.}
\end{figure}

\section{superconductivity from hole undressing}

As discussed in II and Ref. \cite{ijmp}, there is plenty of experimental evidence that dressed $hole$ carriers in the normal state become undressed
from the electron-ion interaction and behave like undressed 
$electron$ carriers in the superconducting state.   How this occurs
becomes  clear from the considerations in this paper.

Let us review the experimental evidence for undressing of  carriers from the electron-ion interaction in the superconducting state:

{\it (1) Rotating superconductor}:  A superconducting body rotating with angular velocity $\vec{\omega}$ develops a uniform magnetic field in the interior\cite{rotating},
given by 
\beq
\vec{B}=-\frac{2 m_e c}{e}\vec{\omega}
\eeq
with $m_e$ the {\it free electron mass}. This has been measured in conventional, heavy fermion, and
high $T_c$ cuprate superconductors\cite{rotating2}. The magnetic field always points $parallel$ to the angular velocity, as given by
Eq. (40), never antiparallel, indicating that the superfluid carriers behave as   {\it negatively charged carriers}\cite{ehas}.

{\it (2) Gyromagnetic effect}:  If a magnetic field $\vec{B}$ is suddenly applied to a superconductor at rest, the entire body is found to start rotating with angular momentum
\beq
\vec{L}_{body}=\frac{m_e c}{2\pi e} V\vec{B}
\eeq
with $m_e$ the free electron mass, and $V$ the volume\cite{gyro}. This angular momentum is always $antiparallel$ to the applied magnetic field for the same
reason as in {\it (1)} above\cite{ehas}.

{\it (3) Bernoulli potential}: 
Where there is a spatial variation of the superfluid velocity an electric field is expected to develop, the resulting potential is termed Bernoulli potential\cite{london}.
Experimental measurements\cite{bernoulli} are consistent with an electric field given by
\beq
\vec{E}=\frac{1}{e}\vec{\nabla}\frac{1}{2} m_e v_s^2
\eeq
where $e$ and $m_e$ are the {\it free electron} charge and mass, and $v_s$ is the superfluid velocity. In particular the $sign$ of the Bernoulli potential measured
corresponds to $negative$ charge carriers.

{\it (4) Hall effect}: The Hall coefficient is found to be essentially always positive in the normal state, corresponding to hole carriers\cite{hallns}, and to change its sign from
positive to negative at temperatures slightly below $T_c$, indicating that carriers change from hole-like to electron-like\cite{hall}.

{\it (5) Wavelength expansion}: We have proposed an explanation for the Meissner effect in superconductors involving a {\it wavelength expansion} of the carriers at
the Fermi energy\cite{hole2,sm} from the
microscopic length $k_F^{-1}$ to the mesoscopic length $2\lambda_L$ ($\lambda_L=$London penetration depth), corresponding 
to the wavevector change $k_F \rightarrow q_0=1/(2\lambda_L)$.

These  experimental observations and theoretical considerations indicate   that the carriers of electric current in the normal state, dressed hole carriers,
 morph into undressed electron-like carriers in the superconducting state. 
How does this happen? In II and Ref.\cite{ijmp}, we suggested that  the antibonding electrons at the top of the Fermi distribution
in the normal state condense to the bottom of the band in the transition to superconductivity. However, this would conflict with the Pauli principle,
because the states at the bottom of the band are occupied!

Furthermore, it is important to point out that it would be incorrect to assume that  $all$ the electrons in the band  become `undressed' from the electron-ion
interaction in the superconducting state. Because the superfluid weight $n_s$ that enters into the London penetration depth $\lambda_L$
\beq
\frac{1}{\lambda_L^2}=\frac{4\pi n_s e^2}{m_e c^2}
\eeq
most definitely corresponds to the $hole$ concentration rather than to the electron concentration in the band (which is much larger). It is essentially the same charge carrier density
that carries the current in the normal state.  

So the $holes$ in the normal state have to behave $like$ $electrons$ in the superconducting state. The solution of this puzzle is 
what was depicted in Fig. 6. In Fig. 9 we show the expected evolution with temperature of the occupations in the band structure and
in the Brillouin zone.
As the metal is cooled into the superconducting state, 
the holes condense to the bottom of the band, which corresponds to long-wavelength states near the center of the Brilloin zone,
which are   smooth bonding states as depicted on the left side of Fig. 7.
So it may be said that the superfluid carriers are still holes, not electrons.  However, {\it the sign
of the effective mass for carriers at the bottom of the band is opposite to that of carriers  at the top of the band}. For that reason, these superfluid  
 hole carriers behave like electrons,  and in particular exhibit a negative Hall coefficient, as well as the other signatures of
 electron transport reviewed at the beginning of this section.

  \begin{figure}
\resizebox{8.5cm}{!}{\includegraphics[width=7cm]{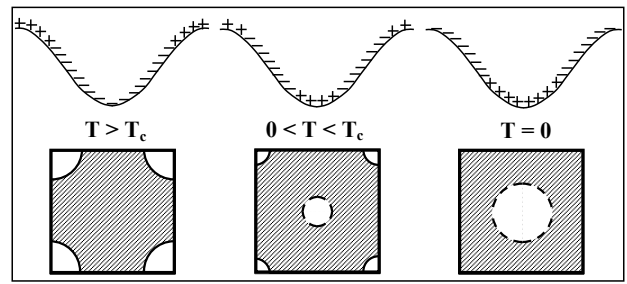}}
  \caption{Occupied states in the band (upper part of the figure) and in the Brillouin zone (lower part of the figure)
   as the temperature is lowered.  The shaded areas of the Brillouin zone are occupied by electrons. When the temperature is lowered  below $T_c$ a `hole pocket' is pierced at $k\sim0$ that grows
  as the temperature is further lowered, while the hole pockets near the edges of the Brillouin zone become progressively smaller and eventually disappear at $T=0$. At temperatures 
  $0<T<T_c$ the empty states near $k \sim \pi$ correspond to the normal fluid hole carriers and the empty states near $k \sim 0$ correspond to  the superfluid hole carriers.}
\end{figure}

Figure 9 also suggests a new interpretation of the `two-fluid model'\cite{twofluid} of superconductivity. The normal quasiparticles are the holes near the top of the band, i.e. $k\sim \pi$, while
the superfluid are the holes at the bottom of the band, with  $k \sim 0$. The conventional BCS theory is argued to be consistent with the two-fluid model\cite{bardeentwofluid}, however it does not
provide a clear separation of both components as Fig. 9 does.

Note also that the continuous process by which the holes move from the top to the bottom of the band as the superfluid condensate
develops represents the momentum space counterpart of
the orbit expansion discussed in connection with the Spin Meissner effect\cite{sm}.

\section{Two-orbital model}
In previous work we have introduced an electronic model with two orbitals per site to describe the essential physics of electron-hole asymmetry
(electronic dynamic Hubbard model)\cite{dynh2}.
The spacing between single-particle energy levels is $\epsilon$, and the interactions are such that two electrons will occupy predominantly the higher single-particle level 
because the Coulomb repulsion there ($U'$) is much smaller than both in the lower level and between one electron in each level. This corresponds to the situation
depicted schematically on the right side of Fig. 1.

  \begin{figure}
\resizebox{8.5cm}{!}{\includegraphics[width=7cm]{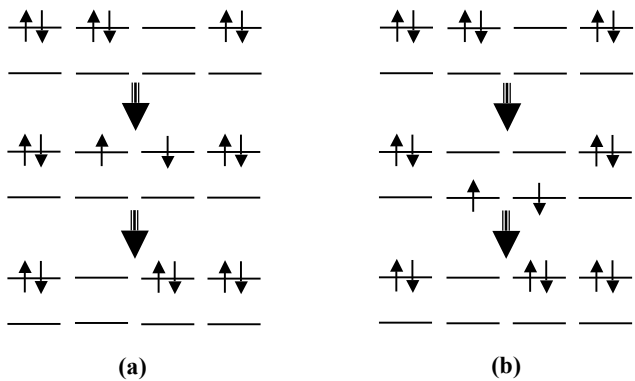}}
  \caption{Propagation of a pair in the strong coupling limit. The fact that the holes undress completely and propagate as free electrons implies there is no relaxation of the charge cloud as depicted
  in (b), rather the propagation is as depicted in (a), with the holes occupying the lower atomic energy level.}
\end{figure}
\

 Consider a pair of holes propagating within this model in the  limit where the size of the pair is a single lattice spacing. This corresponds to the
 strong coupling limit where the single hole hopping amplitude goes to zero as discussed in Ref.\cite{strong}. In Fig. 10 we show two possible ways the pair propagation can occur.
As in Ref.\cite{dynh2} (Eq. (4)), we assume for simplicity that the bare hopping amplitude is the same between all nearest neighbor orbitals, $t$. 
 If the propagation is as shown in Fig. 10(b), the electrons in the singly occupied site relax  to the lower orbital, and the 
 hopping amplitude is reduced by the overlap matrix element $S$ (Eq. 27(a) of Ref.\cite{dynh2}). Instead, if the propagation is as shown in
 Fig. 10(a), the hopping amplitude is not reduced. This corresponds to the case where the $holes$ occupy the lowest orbital, as depicted on the right side of Fig. 6. 
 
 The wavefunction for a pair of holes is a linear combination of the states
 \bmath
 \beq
 |\Psi_p>=\frac{1}{\sqrt{N}}\sum_i|\uparrow\downarrow>_i
 \eeq
 \beq
 |\Psi_{ex}>=\sqrt{\frac{2}{zN}}\sum_{<ij>}[|\uparrow>_i|\downarrow>_j-|\downarrow>_i|\uparrow>_j]
 \eeq
 \emath
 where $i,j$ are nearest neighbor sites and $z$ is the number of nearest neighbors to a site. Its energy is the lowest eigenvalue of the matrix
 \[H_1 = \left| \begin{array}{cc}
0 & -2t\sqrt{z}  \\
-2t\sqrt{z}& -U'   \end{array} \right|\] 
 or the matrix
  \[H_2 = \left| \begin{array}{cc}
0 & -2tS\sqrt{z}  \\
-2tS\sqrt{z}& -U' -2\epsilon  \end{array} \right|\] 
 for the case of Fig. 10(a) and 10(b) respectively. In the limit where $S$ is very small, the condition on the parameters for the lowest eigenvalue of $H_1$ to be smaller than the
 lowest eigenvalue of $H_2$ is 
 \beq
 \sqrt{z}t>\sqrt{\frac{\epsilon U'}{2}+\epsilon^2}
 \eeq
Here, $t$ is the bare electron hopping amplitude for an empty band. For a cubic lattice it is given by
\beq
t=\frac{\hbar^2}{2m_e a^2}=\frac{3.81 eV}{a(\AA)^2}
\eeq
Therefore, the condition Eq. (45) can be satisfied for reasonable parameters, e.g. $a=2\AA$, $t=0.95 eV$, $\epsilon=1eV$, $U'=5eV$.
In that case, the hole propagates without disturbing the background, as shown schematically in the left diagram of Fig. 7.
The propagation shown in Fig. 10(b) corresponds to `partial' undressing from the electron-electron interaction, while that shown in Fig. 10(a) corresponds to full undressing

\section{discussion}
In this paper we have discussed the point of view that the   competition between electron-ion interaction dominance versus electron-electron interaction
dominance in solids leads to normal metallic or superconducting behavior depending on whether the former or the latter wins.
Electron-electron interaction will dominate when an  electronic energy band is almost full, i.e. when the carriers in the normal state are hole-like.
Correspondingly, superconductors in nature are found to have positive Hall coefficient in the normal state\cite{hallns}. Furthermore, the electron-electron interaction strength
$e^2$ will dominate over the electron-ion interaction strength $Ze^2$  when the ionic charge $Z$ is small. Correspondingly, high $T_c$ superconductivity
is found in materials with   highly negatively charged substructures (planes) containing negative ions, like the $(Cu-O_2)^=$  planes in the cuprates, the $(Fe-As)^-$
 planes in the arsenides  and the $B^- $ planes in $MgB_2$.
 
Here we restricted the discussion of the atom and the diatomic molecule  to the simplest case of a $1s$ orbital. However exactly the same physics
 should take place in bands originating from other atomic orbitals. For example we showed in Ref.\cite{diat3} that the quantities of interest for 
 $2p$ orbitals in a diatomic molecule behave very similarly to those in  the $1s$ orbital. With $p$ orbitals, the roles of $k\sim 0$ and $k\sim \pi$ switch, however the essential physics
 determined by whether a state is near the bottom or near the top of the band remains the same.
 
 When one talks about electronic energy bands one is implicitly privileging the electron-ion interaction over the electron-electron interaction. The states at the bottom of a band have low
 electron-ion energy, and those at the top of a band have high electron-ion energy. However, conversely, as we have argued in this paper, electrons residing in states near the 
 bottom of the band have  high electron-electron repulsion energy, and those near the top of the band have  low electron-electron repulsion energy. It is only natural to
 assume that in certain cases a complete reorganization can  occur and a new state will emerge that optimizes the electron-electron interaction instead of  the electron-ion
 interaction. That, we propose, is the superconducting state: electrons redistribute their occupation in the band states to occupy the high electron-ion energy states, low electron-electron
 energy states, and
 leave empty the low electron-ion energy states, high electron-electron energy states. The concept of  ``holes'' provides a natural language to describe the process: holes condense to occupy the low
 electron-ion states, giving rise to hole superconductivity.
 
The physics proposed in this paper also illustrates an even closer connection between the phenomena of superconductivity and metallic ferromagnetism than previously suspected\cite{hydro}. 
 In our previous work we proposed that both phenomena
 originate in electronic ``bond charge repulsion''\cite{bondcharge,bondchargefm,bondchargefm2}, both lead to lowering of the carrier effective mass\cite{apparent,mstarfm} 
 and ``undressing''\cite{undr,undrfm}, and both are  driven by off-diagonal matrix elements of the Coulomb interaction\cite{fmscmstar}, with
 ferromagnetism dominating near the half-filled band\cite{fm} and superconductivity when the band is almost full. For metallic ferromagnetism it was always clear that the phenomenon involves
 occupying some states that are unfavorable for the electron-ion interaction and emptying some states that are favorable to the electron-ion interaction. Namely,
  the majority spin electrons occupy antibonding states that were empty in the non-ferromagnetic state, and some
 bonding states that were occupied by the minority spins in the normal state become empty in the ferromagnetic state\cite{fm}. With the interpretation of superconductivity proposed in this paper
 it is clear that something quite similar occurs in the superconducting state, with empty antibonding states becoming full and full bonding states becoming empty.
  
 In previous work we have emphasized the contribution to $\Delta t$ arising solely from modulation of the single particle hopping amplitude by the on-site orbital expansion overlap matrix elements
 (Frank-Condon factor)\cite{hole1}. That physics is contained in the `dynamic Hubbard model'\cite{dynh} with an on-site Coulomb repulsion modulated by a local boson degree of freedom,
 or in a purely electronic model with two orbitals per site and only on-site interactions\cite{dynh2}. These models have much of the relevant physics: they give rise to pairing and
 superconductivity driven by ``undressing'' from the electron-electron interaction\cite{dynh3,dynh4}. However these models don't  allow for a change in sign of the single particle hopping amplitude, thus  will not
 lead to  ``undressing'' from the electron-ion interaction. That physics requires inclusion of off-site Coulomb matrix elements in the
 Hamiltonian as discussed in Sect. III.
  
In summary, in this paper we have proposed a new physical picture to describe superconductors: that it is the $holes$ that condense to the bottom of the electronic energy band when a system becomes
superconducting. This naturally   ties together several elements introduced earlier within the theory of hole superconductivity: it explains why carriers undress from the electron-electron and the
electron-ion interactions and behave as completely free electrons, yet their $number$ is the number of holes rather than the number of electrons in the band; the
physics is tied to the Coulomb matrix element $\Delta t$ that has played a key role since the beginnings of this theory\cite{bondcharge}; finally,  we have shown that this physics is connected to the negative
charge expulsion from the interior of superconductors previously found based on different arguments\cite{chargeexp}.

In a finite cluster, this reorganization of energy level occupation should also occur, amongst the discrete energy levels of the cluster.
It has not escaped our notice that this may provide an explanation for the remarkable experimental observations of de Heer and coworkers\cite{deheer}.

 Further discussion and development of this physics and its connection with other elements of the theory will be given in future work.

 \end{document}